\documentclass[aps,pra,amsmath,amssymb,twocolumn,superscriptaddress,showpacs]{revtex4-1}
\usepackage{amsmath,amssymb}
\usepackage{graphicx}	
\usepackage{bm}
\usepackage{graphicx}	

\def\vecb{\boldsymbol}
\def\be{\begin{eqnarray}}
\def\ee{\end{eqnarray}}

\begin{document}

\author{Shunsuke~A.~Sato}
\email{ssato@ccs.tsukuba.ac.jp}
\affiliation 
{Center for Computational Sciences, University of Tsukuba, Tsukuba 305-8577, Japan}
\affiliation 
{Max Planck Institute for the Structure and Dynamics of Matter, Luruper Chaussee 149, 22761 Hamburg, Germany}

\title{First-principles calculations for transient absorption of laser-excited magnetic materials}

\begin{abstract}
We investigate the modification in the optical properties of laser-excited bulk cobalt and nickel using the time-dependent density functional theory at a finite electron temperature. As a result of the first-principles simulation, a complex change in the photoabsorption of the magnetic materials is observed around the $M_{2,3}$ absorption edge. Based on the microscopic analysis, we clarify that this complex absorption change consists of the two following components: (i) the decrease in the photoabsorption in a narrow energy range around the $M_{2,3}$ edge, which reflects the blue shift of the absorption edge due to the light-induced demagnetization, and (ii) the increase in the photoabsorption in a wider range around the $M_{2,3}$ edge, which reflects the modification in the local-field effect due to the light-induced electron localization. The relation between the transient optical and magnetic properties may open a way to monitor ultrafast (de)magnetization and spin dynamics in magnetic materials via transient absorption spectroscopy.
\end{abstract}

\maketitle

\section{Introduction \label{sec:intro}}

Owing to the recent developments in laser technologies, the study of highly nonlinear nonequilibrium electron dynamics in solids has become experimentally accessible \cite{RevModPhys.72.545,RevModPhys.81.163,MOUROU2012720}, offering novel opportunities for investigating the optical control of electric transport \cite{Schiffrin2013,Higuchi2017,PhysRevB.99.214302,McIver2020} and the emergence of a dynamical phase of matter \cite{PhysRevB.79.081406,Lindner2011}. In the ultrafast regime, the transient optical properties of solids that are driven by intense light pulses have been intensively studied in the attosecond regime, and a microscopic understanding of the nonequilibrium electron dynamics during and after laser excitation has been obtained \cite{doi:10.1126/science.1260311,Zurch2017,doi:10.1126/science.aag1268,doi:10.1126/science.aan4737,doi:10.1063/5.0020649,Lucchini2021}. Recently, the attosecond circular dichroism technique, which enables the study of light-induced ultrafast spin dynamics in magnetic materials, has been developed, opening a novel avenue toward achieving optical control of ultrafast spin dynamics and magnetization \cite{Siegrist2019}.

First-principles electron dynamics simulations based on the time-dependent density functional theory (TDDFT) have been a powerful tool to obtain microscopic insight into light-induced ultrafast phenomena \cite{PhysRevLett.52.997,PhysRevB.62.7998}. Recently, this approach has been used in the field of attosecond transient absorption spectroscopy in solids, enabling access to the microscopic mechanism behind the macroscopic ultrafast phenomena \cite{doi:10.1126/science.aag1268,Volkov2019,Siegrist2019}.

In this work, we study the transient optical properties of laser-excited magnetic materials, namely cobalt and nickel, using first-principles electron dynamics simulations, aiming to find a signature of light-induced magnetization and spin dynamics in the transient absorption spectra. Once the relation between magnetization dynamics and transient absorption is established, a complementary and alternative technique to the attosecond circular dichroism technique is established to investigate light-induced magnetization and spin dynamics. Through microscopic simulations, we find a complex modification in the photoabsorption of the laser-excited magnetic materials around their $M_{2,3}$ absorption edge. This modification reflects light-induced electron localization around the ions and a reduction in the a exchange splitting in the $3p$ semicore bands. Hence, the microscopic information of magnetic systems is indeed integrated in the transient absorption spectrum.

This paper is organized as follows. In Sec.~\ref{sec:method}, we first revisit the electron dynamics simulations with the TDDFT and describe a method to compute the optical properties of solids. In Sec.~\ref{sec:result}, we study the transient optical properties of laser-excited magnetic materials and analyze the microscopic origin behind the observed modification of photoabsorption. Finally, our findings are summarized in Sec.~\ref{sec:summary}. Atomic units are used unless stated otherwise.

\section{Methods \label{sec:method}}

\subsection{Electron dynamics simulation \label{subsec:elec-dynamics}}

Firstly, we briefly revisit the electron dynamics simulations based on the TDDFT \cite{PhysRevLett.52.997,PhysRevB.62.7998}. The details of the method are described elsewhere \cite{SATO2021110274}. The light-induced electron dynamics in solids is described by the following time-dependent Kohn--Sham equation,
\be
i\frac{\partial}{\partial t}u_{b,s,\vecb{k}}(\vecb r,t)=\hat h_{s,\vecb k}(t)u_{b,s,\vecb{k}}(\vecb r,t),
\label{eq:tdks}
\ee
where $b$ is the band index, $s$ is the spin index ($s=\uparrow$ or $\downarrow$), $\vecb k$ is the Bloch wavevector, and $u_{b,s,\vecb k}(\vecb r,t)$ represents the periodic part of the time-dependent Bloch orbitals. The time-dependent Kohn--Sham Hamiltonian, $\hat h_{s,\vecb k}(t)$, is given by
\be
\hat h_{s,\vecb k}(t) = \frac{\left [ \vecb p + \vecb k + \vecb A(t) \right ]^2}{2}
+ \hat v_{ion} + v_{Hxc}\left [ \rho_e(\vecb, t), \rho_s(\vecb r,t) \right ], \nonumber \\
\label{eq:ks-ham}
\ee
where $\vecb A(t)$ is a time-varying spatially-uniform vector potential, which is related to the external laser electric field as $\vecb E(t)=-\dot{\vecb A}(t)$. The ionic potential is denoted as $\hat v_{ion}$, and it may consist of spatially non-local parts when using the norm-conserving pseudopotential method \cite{PhysRevLett.48.1425,PhysRevB.43.1993}. The Hartree-exchange-correlation potential is denoted as $v_{Hxc}\left [\rho_e(\vecb r,t),\rho_s(\vecb r,t) \right ]$, and it is a functional of the electron density, $\rho_e(\vecb r,t)$, and the spin density, $\rho_s(\vecb r,t)$. The total electron density, $\rho_e(\vecb r,t)$, is given by the sum of the spin density as $\rho_e(\vecb r,t)=\rho_{\uparrow}(\vecb r,t) + \rho_{\downarrow}(\vecb r,t)$, and each spin density is defined as
\be
\rho_s(\vecb r,t) = \sum_{b} \frac{1}{\Omega_{BZ}}\int_{BZ} d\vecb k f_{b,s,\vecb k}\left |u_{b,s,\vecb k}(\vecb r,t) \right|^2,
\label{eq:rho}
\ee
where $\Omega_{BZ}$ is the volume of the Brillouin zone, and $f_{b,s,\vecb k}$ is the occupation of the orbital. We define the occupation factor as a Fermi--Dirac distribution, that is
\be
f_{b,s,\vecb k} = \frac{1}{e^{(\epsilon_{b,s,\vecb k}-\mu)/k_BT_e}+1},
\ee
where $\mu$ is the chemical potential, $T_e$ is the electron temperature, and $\epsilon_{b,s,\vecb k}$ are the single-particle energies of the self-consistent solutions of the following static Kohn--Sham equations at $t=-\infty$:
\be
\hat h_{s,\vecb k}(t=-\infty)u_{b,s,\vecb{k}}(\vecb r,t=-\infty) = \epsilon_{b,s,\vecb{k}} u_{b,s,\vecb{k}}(\vecb r,t=-\infty). \nonumber \\
\ee
Note that the eigenstates $u_{b,s,\vecb{k}}(\vecb r,t=-\infty)$ are used as the initial conditions for Eq.~(\ref{eq:tdks}).

\begin{widetext}
Using these time-dependent Kohn--Sham orbitals, $u_{b,s,\vecb{k}}(\vecb r,t)$, one can evaluate the electric current as a function of time as
\be
\vecb J(t) = - \frac{1}{\Omega_{cell}} \sum_{b,s} \frac{1}{\Omega_{BZ}}\int_{BZ} d \vecb k 
f_{b,s,\vecb k} \int_{cell} d\vec r u^*_{b,s,\vecb k}(\vecb r,t) \vecb v_{\vecb k}(t) u_{b,s,\vecb k}(\vecb r,t),
\ee
where $\Omega_{cell}$ is the volume of the unit cell, and $\vecb v_{\vecb k}(t)$ is the velocity operator defined as
\be
\vecb v_{\vecb k}(t) = \frac{\left [\vecb r, \hat h_{s,\vecb k}(t) \right ]}{i\hbar},
\ee
where the commutator means $\left [A,B \right ]=AB-BA$. In this work, we evaluate the optical properties of materials using the computed current, $\vecb J(t)$.
\end{widetext}

\subsection{Linear response calculation for the optical absorption of solids \label{subsec:lin-res}}

As described in Sec.~\ref{subsec:elec-dynamics}, one can evaluate the electric current induced by given electric fields. Hence, the TDDFT simulation yields the current as a functional of electric fields, namely $\vecb J(t)=\vecb J[\vecb E(t)](t)$. In the weak-field limit, $|\vecb E(t)|\rightarrow 0$, each component of the current can be expanded up to the first order of the electric fields as
\be
J_{\alpha}(t) = \sum_{\beta}\int dt' \sigma_{\alpha \beta}(t-t') E_{\beta}(t'),
\label{eq:lin-res-rt}
\ee
where $J_{\alpha}$ is the $\alpha$-component of $\vecb J(t)$, $E_{\beta}(t)$ is the $\beta$-component of $\vecb E(t)$, and $\sigma_{\alpha \beta}(t)$ is the linear conductivity tensor in the time domain. By applying to the Fourier transformation to Eq.~(\ref{eq:lin-res-rt}), one can obtain the linear optical conductivity in the frequency domain as
\be
\tilde J_{\alpha}(\omega) = \sum_{\beta} \tilde \sigma_{\alpha \beta}(\omega) \tilde E_{\beta}(\omega),
\label{eq:lin-res-fr}
\ee
where $\tilde J_{\alpha}(\omega)$, $\tilde E_{\beta}(\omega)$, and $\tilde \sigma_{\alpha \beta}(\omega)$ are the Fourier transforms of $J_{\alpha}(t)$, $E_{\beta}(t)$, and $\sigma_{\alpha \beta}(t)$, respectively. Therefore, by analyzing the applied electric fields and induced current, one can evaluate the linear optical conductivity, $\tilde \sigma_{\alpha \beta}(\omega)$. Furthermore, one can evaluate the dielectric function as
\be
\epsilon_{\alpha \beta} = \delta_{\alpha \beta} + 4\pi i \frac{\tilde \sigma_{\alpha \beta}}{\omega}
\label{eq:eps}
\ee
and the absorption coefficient as
\be
\mu_{\alpha}(\omega) = \frac{2\omega}{c} \mathrm{Im}\left [ \sqrt{\epsilon_{\alpha \alpha}} \right ].
\label{eq:mu}
\ee

For practical calculations, we employ an impulsive distortion as the electric field that corresponds to the following vector potential:
\be
\vecb A(t) = - k_0 \vecb e_{\vecb \beta} \Theta(t),
\label{eq:impulsive-kick}
\ee
where $k_0$ is strength of the impulsive distortion, $\vecb e_{\beta}$ is the unit vector along the $\beta$-direction, and $\Theta(t)$ is the Heaviside step function. Under the impulsive distortion described by Eq.~(\ref{eq:impulsive-kick}), we solve Eq.~(\ref{eq:tdks}) and compute the induced electric current $\vecb J(t)$. Based on Eq.~(\ref{eq:lin-res-fr}), we then evaluate the optical conductivity as
\be
\tilde \sigma_{\alpha \beta}(\omega) = \frac{1}{k_0}\int^{\infty}_0dt J_{\alpha}(t)e^{i\omega t -\gamma t},
\ee
where $\gamma$ is an effective damping parameter used to reduce the numerical noise due to the finite time of the simulations. In this work, we set $\gamma$ to $0.5$~eV. Using Eqs.~(\ref{eq:eps}) and (\ref{eq:mu}), we evaluate the absorption coefficients of the solids.

In this work, we study the absorption coefficients of bulk cobalt and nickel by changing the electron temperature, $T_e$, in order to analyze the transient optical properties of the materials after laser irradiation \cite{PhysRevB.90.174303}. All calculations in this work were performed using the \textit{Octopus code} \cite{doi:10.1063/1.5142502}. We employ the adiabatic local spin density approximation (ALSDA) for the exchange-correlation functional \cite{PhysRevB.45.13244}. For bulk cobalt, we employ a hexagonal single crystal structure, and we set the lattice parameter $a$ to $2.51$~$\mathrm{\AA}$ and the lattice constant ratio $c/a$ to $1.622$ \cite{hermann2017crystallography}. The hexagonal lattice is discretized into $24\times 24 \times 38$ real-space grid points, and the first Brillouin zone is discretized into $24^3$ $k$-points. For bulk nickel, we employ an $\mathrm{fcc}$ lattice structure, and we set the lattice constant $a$ to $3.52$~$\mathrm{\AA}$ \cite{hermann2017crystallography}. The $\mathrm{fcc}$ lattice is discretized into $28^3$ real-space grid points, and the corresponding Brillouin zone is discretized into $24^3$ $k$-points. Both cobalt and nickel atoms are described using a norm-conserving pseudopotential method, treating $3s$, $3p$, $3d$, and $4s$ electrons as valence electrons \cite{PhysRevB.43.1993,PhysRevB.68.155111,OLIVEIRA2008524}.

\section{Results \label{sec:result}}

\subsection{Electronic structure and optical absorption of magnetic materials \label{subsec:band-absorption}}

To study the optical properties of laser-excited magnetic materials, we first revisit the electronic structures of bulk cobalt and nickel as well as their optical properties in the equilibrium phase.

\begin{figure}[htbp]
  \centering
  \includegraphics[width=0.95\columnwidth]{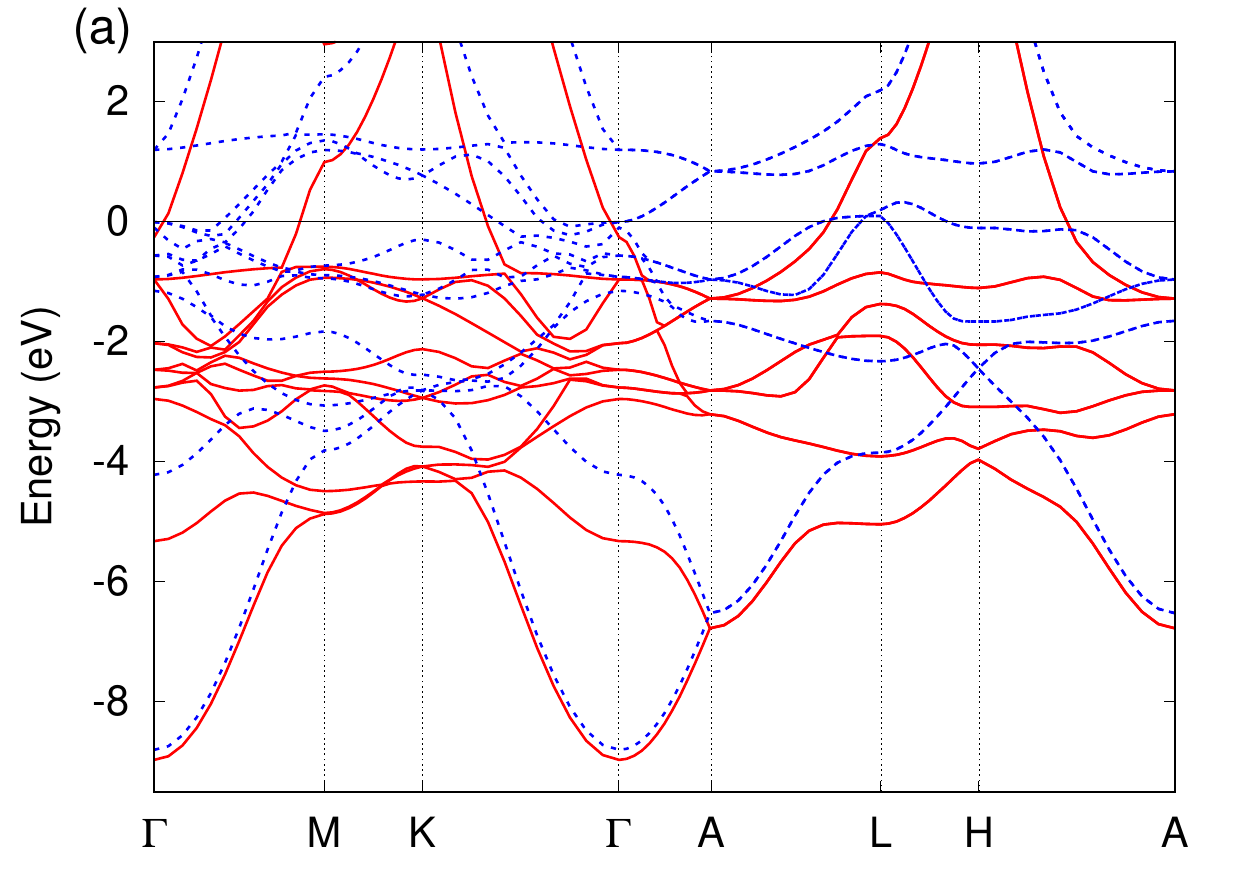}
  \includegraphics[width=0.95\columnwidth]{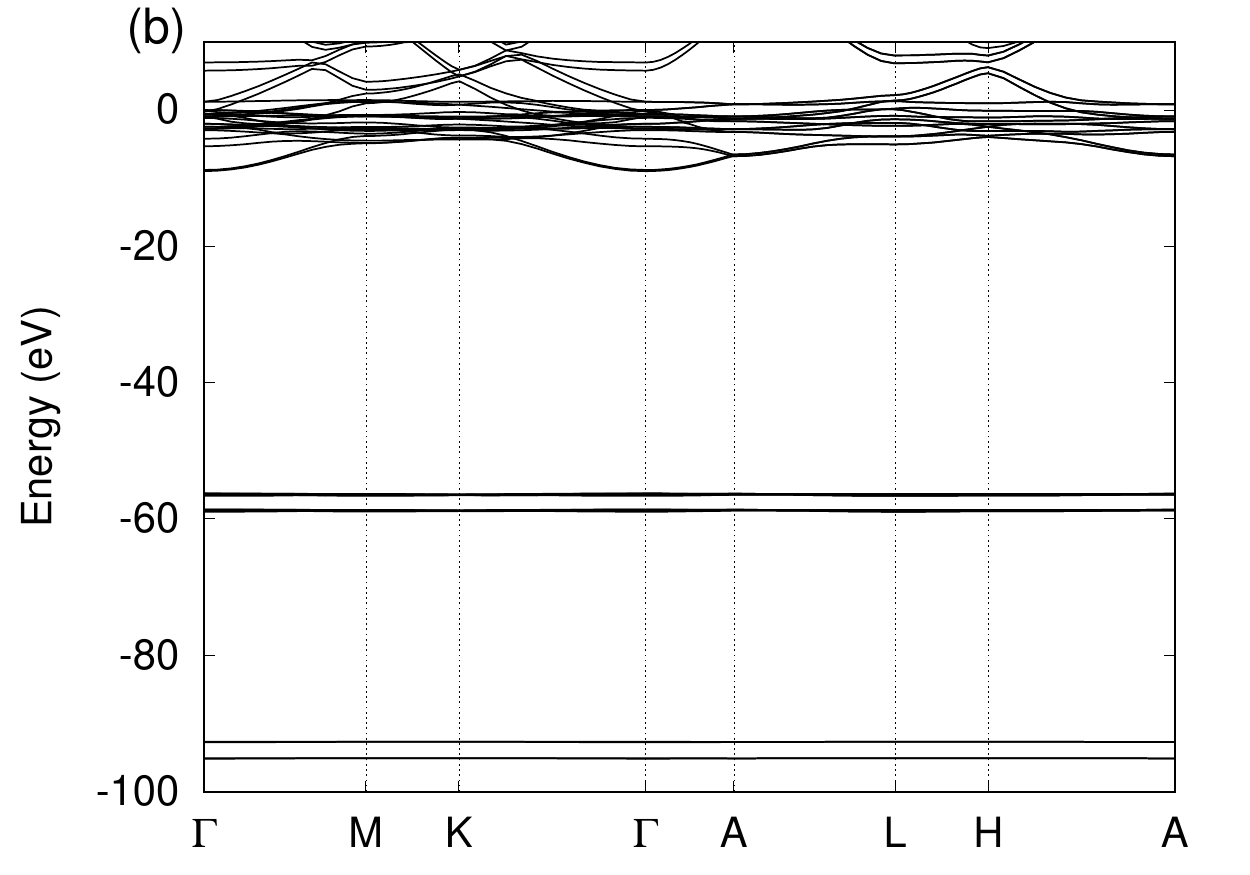}
\caption{\label{fig:Co_hcp_band_wide}
Electronic structure of bulk cobalt. The band structure is computed at $T_e=300$~K. Panel~(a) shows the majority (red) and minority (blue) spin bands around the Fermi surface. Panel~(b) shows the same electronic structure as panel~(a) but in a wider energy range, which includes the Co $3s$ and $3p$ bands.
}
\end{figure}

Figure~\ref{fig:Co_hcp_band_wide}~(a) shows the electronic structure of bulk cobalt. The majority (red-solid line) and minority (blue-dotted line) spin bands are illustrated separately. Here, the electron temperature $T_e$ is set to $300$~K. As seen from the figure, bulk cobalt exhibits the non-degenerate spin bands and is indeed a magnetic system in the present simulation. The computed magnetic moment $\mu$ at $T_e=300$~K is $1.7\mu_B$ per atom. Figure~\ref{fig:Co_hcp_band_wide}~(b) shows the same electronic structure of bulk cobalt but in a wider energy region. Here, the cobalt $3s$ bands are located at around $-95$~eV from the Fermi energy, and the $3p$ bands are located at around $-60$~eV. In both semicore ($3s$ and $3p$) bands, large energy splits are observed. These are exchange splits due to the intrinsic magnetization and are consistent with previously reported large exchange splits \cite{PhysRevLett.104.187401,PhysRevB.89.140404}. The computed exchange splitting of the cobalt $3p$ bands in Fig.~\ref{fig:Co_hcp_band_wide}~(b) is about $2.3$~eV.

\begin{figure}[htbp]
  \centering
  \includegraphics[width=0.95\columnwidth]{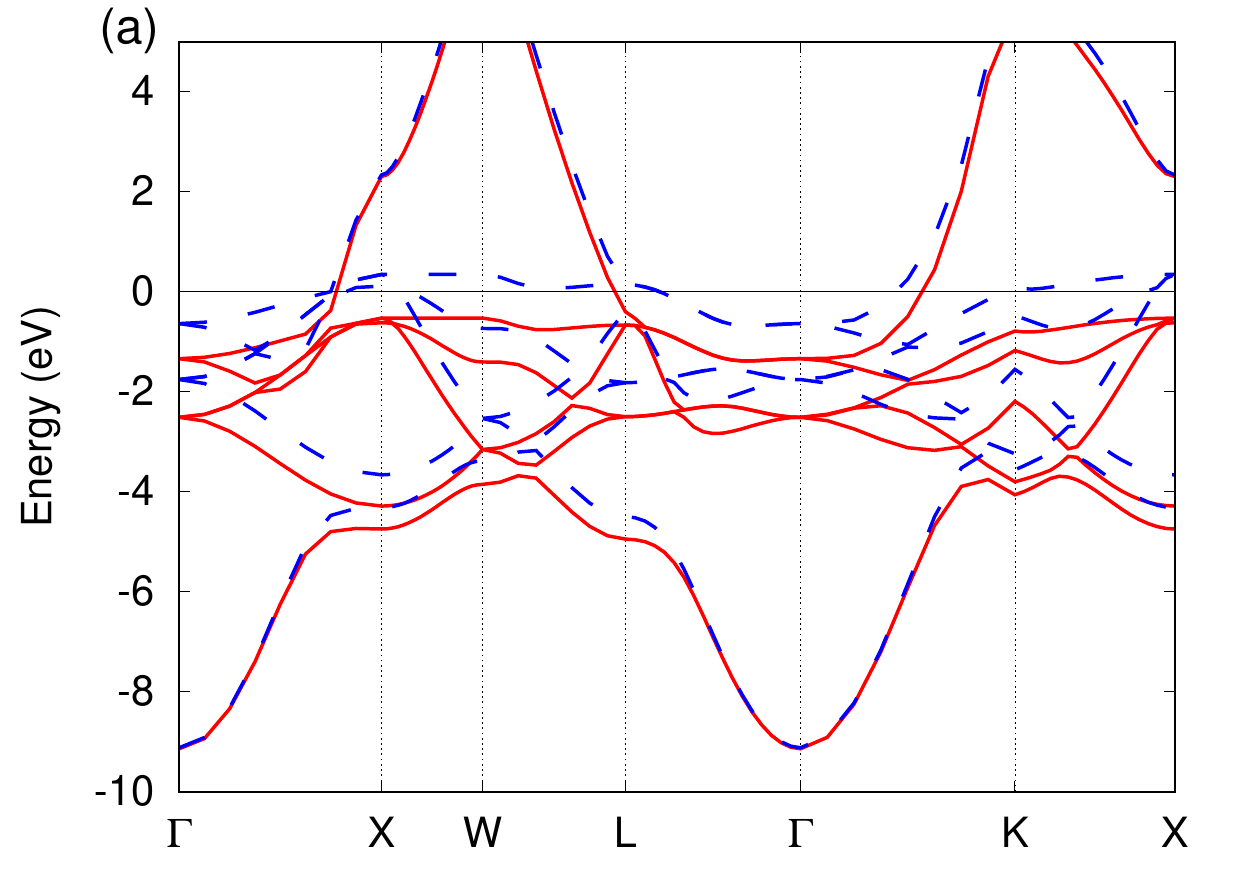}
  \includegraphics[width=0.95\columnwidth]{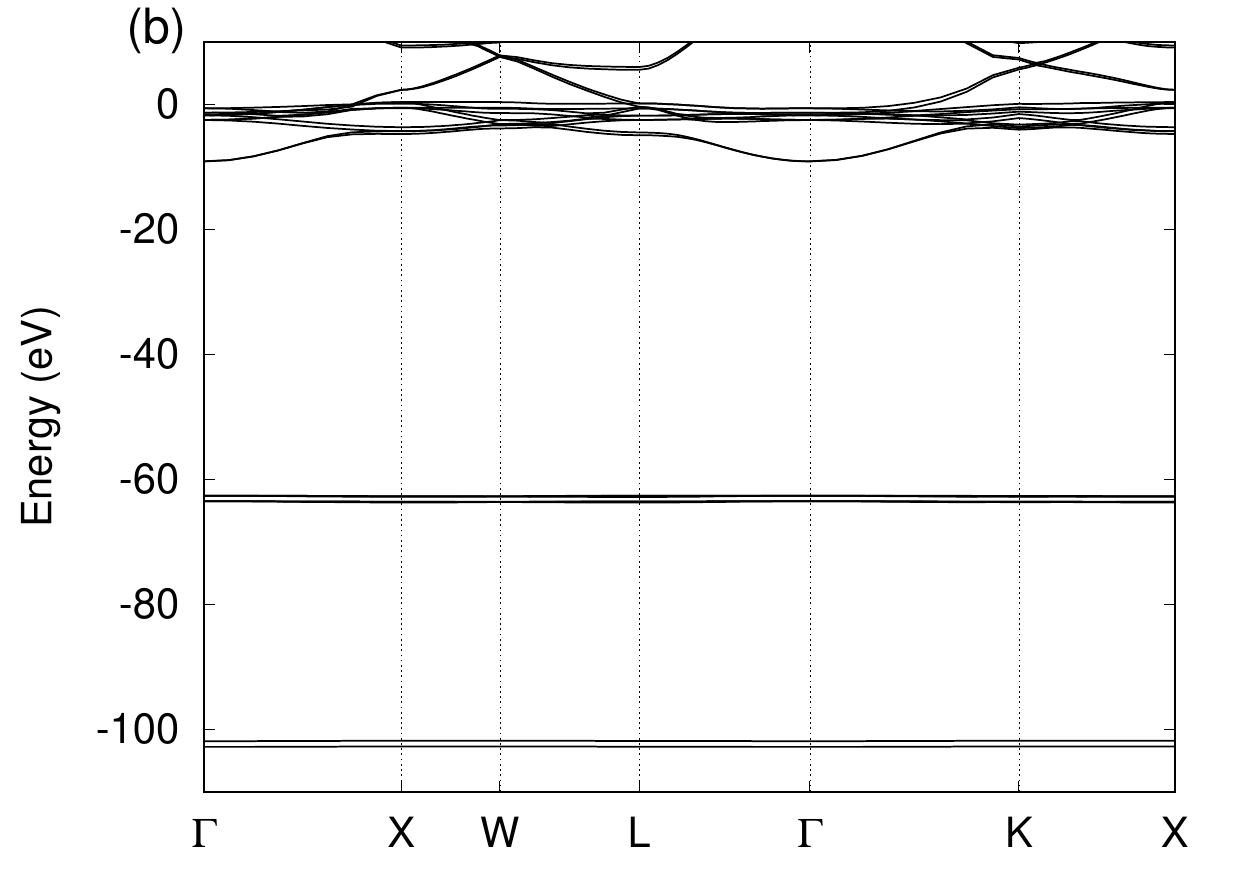}
\caption{\label{fig:band_structure_Ni_wide}
Electronic structure of bulk nickel. The band structure is computed at $T_e=300$~K. Panel~(a) shows majority (red) and minority (blue) spin bands around the Fermi surface. Panel~(b) shows the same electronic structure as panel~(a) but in a wider energy range, which includes the Ni $3s$ and $3p$ bands.
}
\end{figure}

Figure~\ref{fig:band_structure_Ni_wide}~(a) shows the electronic structure of bulk nickel, in which both the majority and minority spin bands are illustrated. In Fig.~\ref{fig:band_structure_Ni_wide}~(b), the same electronic structure is shown for a wider energy range. As for bulk cobalt (Fig.~\ref{fig:Co_hcp_band_wide}), bulk nickel is also a magnetic material, and the two spin bands are non-degenerate. The computed magnetic moment at $T_e=300$~K is $0.67\mu_B$ per atom. As seen from Fig.~\ref{fig:band_structure_Ni_wide}~(b), the semicore bands of nickel also exhibit a large exchange split. The computed exchange split of the nickel $3p$ bands is about $1$~eV.

\begin{figure}[htbp]
  \centering
  \includegraphics[width=0.95\columnwidth]{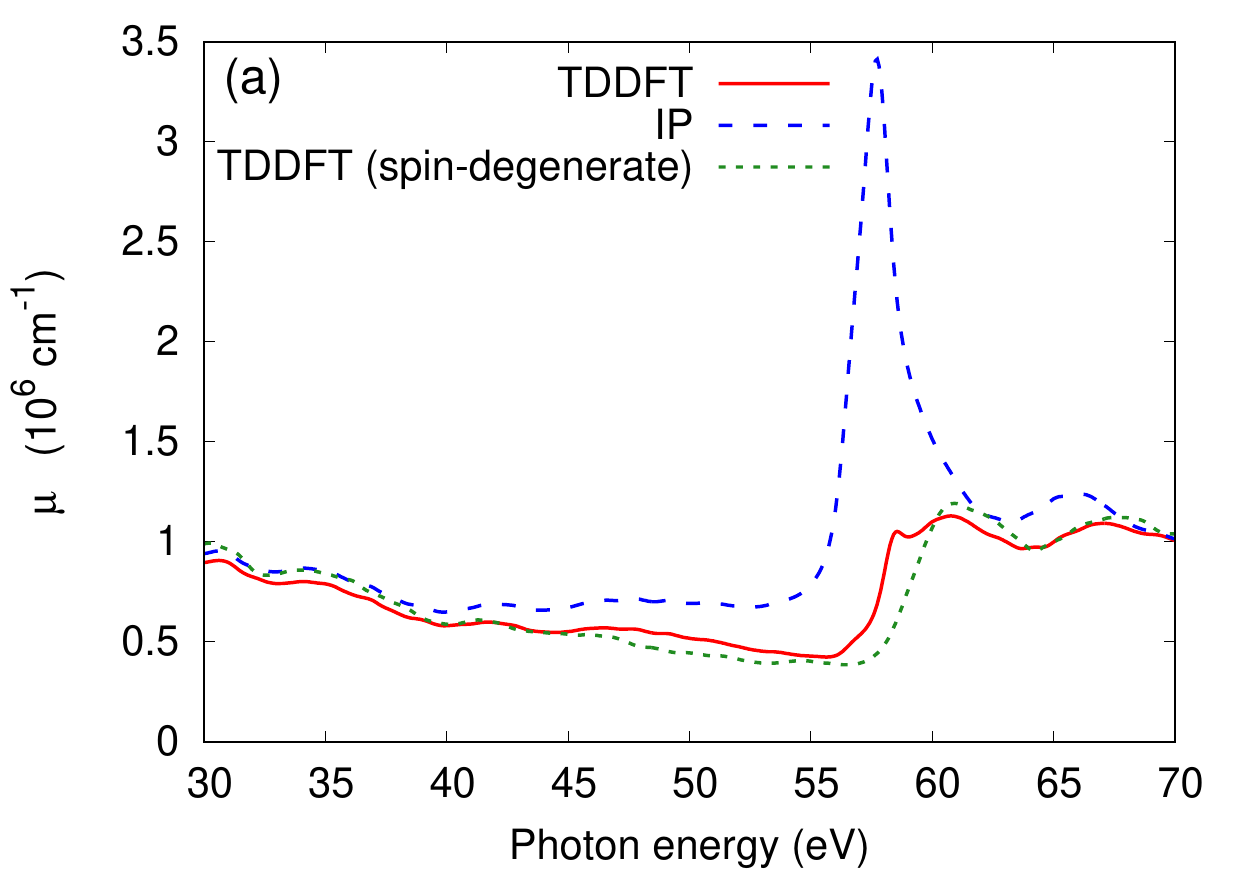}
  \includegraphics[width=0.95\columnwidth]{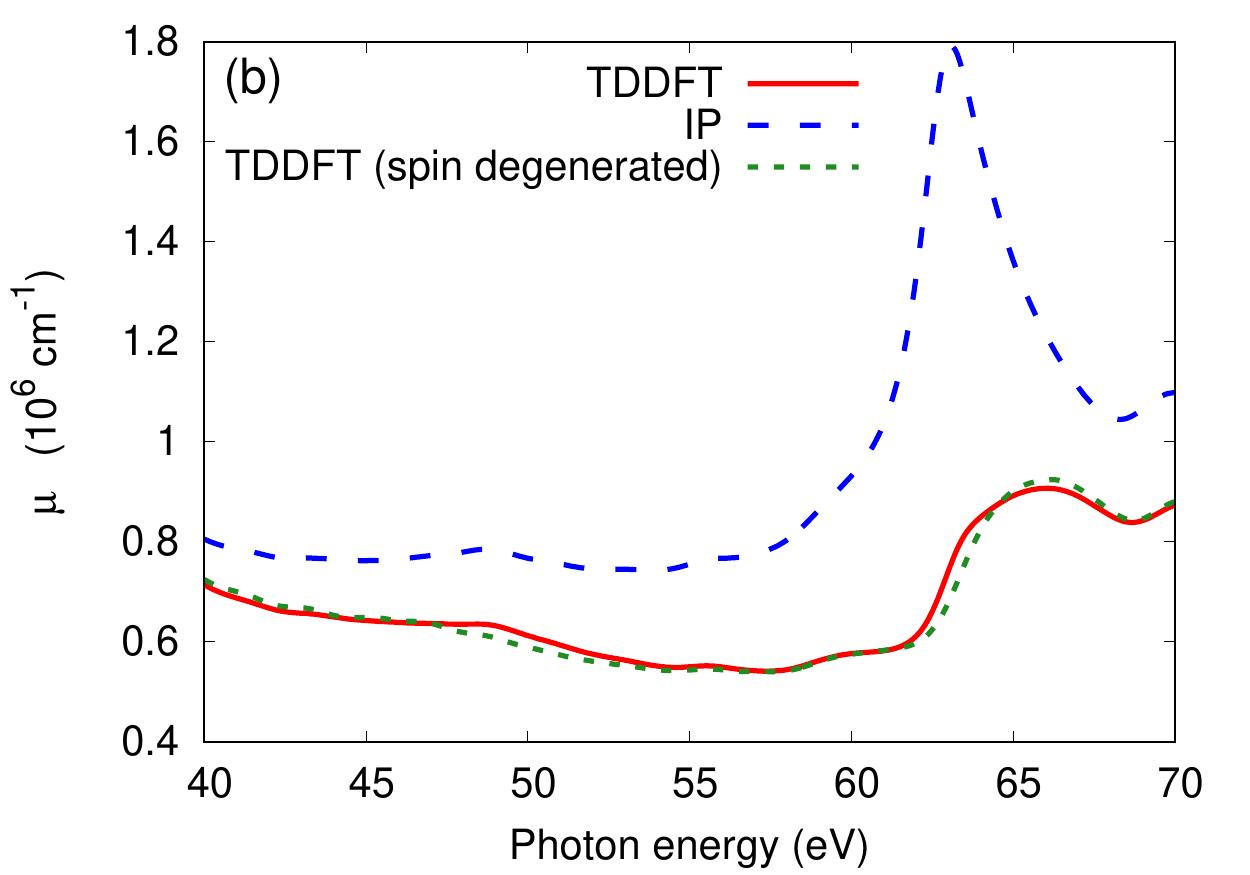}
\caption{\label{fig:mu_Co}
Photoabsorption spectra of (a) cobalt and (b) nickel. In both panels, the results of the full TDDFT simulation are shown by the red solid lines, those of the IP approximation are shown by the blue dashed lines, and those of the TDDFT simulation with the spin-degenerate electronic configuration are shown by the green dotted lines.
}
\end{figure}

Having revisited the basic electronic structure of cobalt and nickel, we investigate the optical properties of these magnetic materials in the equilibrium phase via real-time linear-response calculations based on the TDDFT introduced in Sec.~\ref{subsec:lin-res}. Figure~\ref{fig:mu_Co}~(a) shows the computed absorption coefficient of bulk cobalt with an electron temperature of $300$~K. Here, three simulation results are shown. The red solid line shows the results of the full TDDFT simulation with the ALSDA. The blue dashed line shows the results obtained using the independent-particle (IP) approximation, where the time-dependence of the Hartree-exchange-correlation potential is ignored, and $v_{Hxc}[\rho_e(\vecb r,t), \rho_s(\vecb r,t)]$ is frozen at $t=0$. The green dotted line shows the results of the TDDFT simulation using the spin-degenerate electronic configuration.

As seen from Fig.~\ref{fig:mu_Co}~(a), the full TDDFT result shows a sharp increase in the photoabsorption at around $56$~eV. This energy corresponds to the $M_{2,3}$ absorption edge of bulk cobalt. Around the $M_{2,3}$ edge, the result obtained using the IP approximation shows a large deviation from the full TDDFT simulation result. This indicates that the microscopic screening effect in the time-dependent Hartree potential has a significant role in the photoabsorption around the $M_{2,3}$ edge of cobalt. This is known as the strong local-field effect for semicore electronic responses of transition metals, reflecting a large spatial overlap between the electronic states of the $3p$ and $3d$ bands \cite{PhysRevB.60.R16251}.

From Fig.~\ref{fig:mu_Co}~(a), it can be seen that the result of the spin-degenerate TDDFT simulation is similar to that of the full TDDFT simulation, exhibiting a blue shift of the spectrum around the $M_{2,3}$ absorption edge. The blue shift of the absorption can be understood as the disappearance of the exchange split of the $3p$ bands in the spin-degenerate electronic configuration: In the spin nondegenerate configuration, as shown in Fig.~\ref{fig:Co_hcp_band_wide}~(b), the higher-energy $3p$ band is closer to the Fermi surface than the other spin band due to the exchange split. Once the exchange split vanishes due to the spin degeneracy, the energy difference between the $3p$ bands and the Fermi surface effectively increases, resulting in the blue shift of the absorption edge. This circumstance suggests the possibility to monitor the magnetization of materials via the position of the semicore absorption edge.

Figure~\ref{fig:mu_Co}~(b) shows the absorption coefficient of bulk nickel. By comparing Fig.~\ref{fig:mu_Co}~(a) and Fig.~\ref{fig:mu_Co}~(b), it can be observed that cobalt and nickel show similar behaviors. Therefore, the features discussed above are not unique properties of cobalt but rather general properties of magnetic materials.

\subsection{Transient optical properties of laser-excited magnetic materials \label{subsec:hhg}}

\begin{figure}[htbp]
  \centering
  \includegraphics[width=0.95\columnwidth]{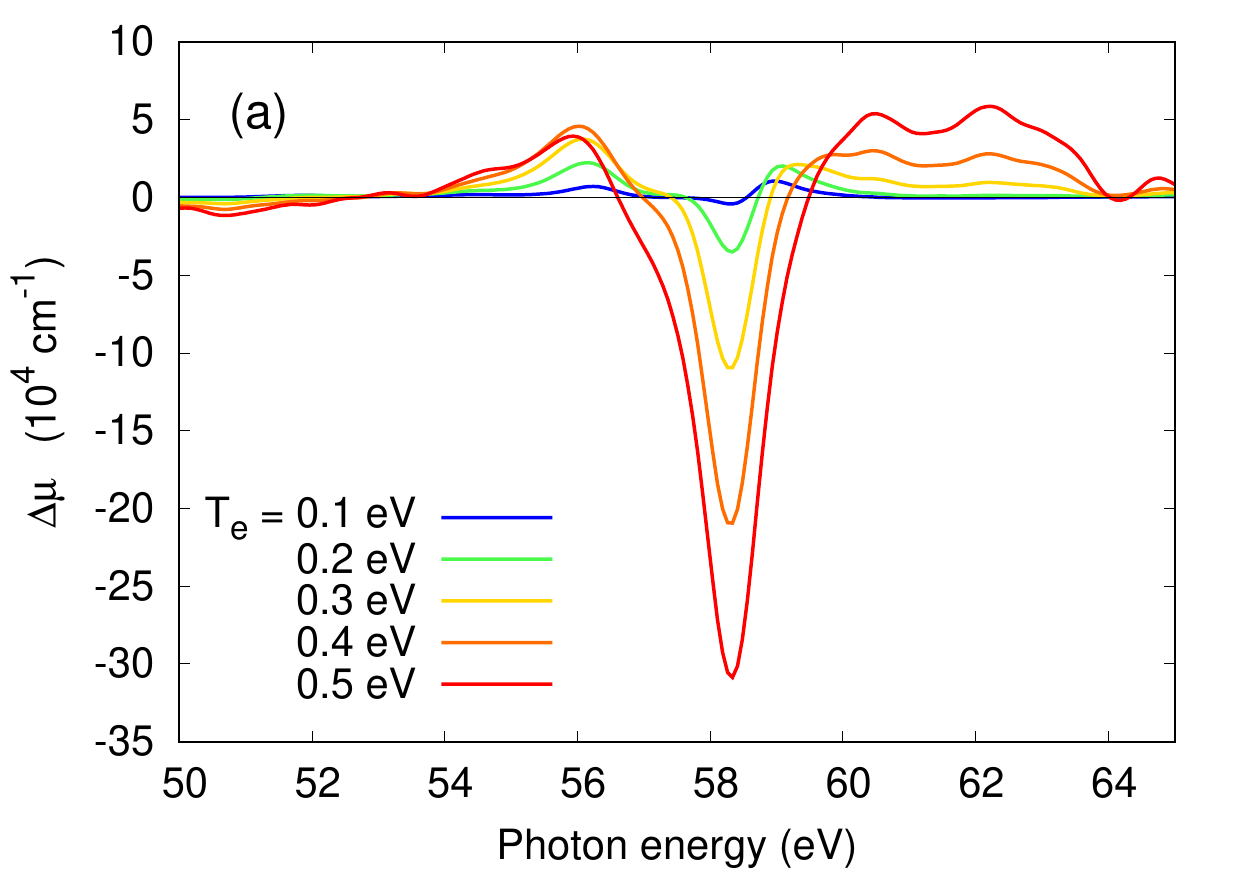}
  \includegraphics[width=0.95\columnwidth]{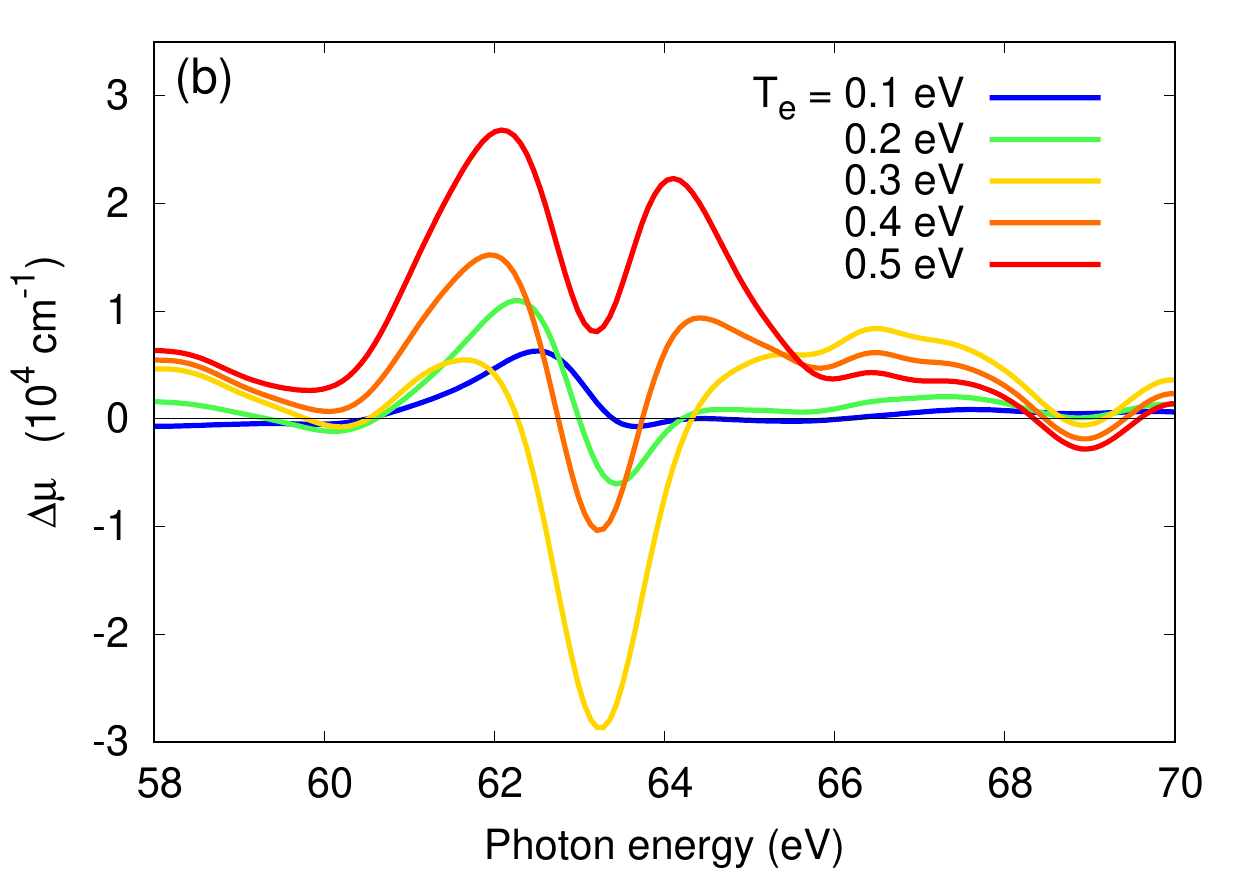}
\caption{\label{fig:delta_mu_Co}
Change in the photoabsorption, $\Delta \mu$, for different electron temperatures, $T_e$, for (a) Co and (b) Ni. The results were obtained using the TDDFT with different electron temperatures.
}
\end{figure}

Here, we study the optical properties of laser-excited magnetic materials using finite-electron-temperature TDDFT simulations by mimicking the two-temperature state of laser-excited matter, in which the electronic system reaches a thermalized hot-electron state while the ionic system remains cold \cite{PhysRevB.90.174303}. We evaluate the change in the absorption coefficient, $\mu(\omega)$, by increasing the electron temperature, $T_e$. Figure~\ref{fig:delta_mu_Co}~(a) shows the change in the absorption coefficient of cobalt for different electron temperatures, $\Delta \mu(\mu,T_e)=\mu(\omega, T_e)-\mu(\omega, T_e=300~\mathrm{K})$, from room temperature ($T_e=300~\mathrm{K}$). Similarly, Figure~\ref{fig:delta_mu_Co}~(b) shows the results for bulk nickel. As seen from Fig.~\ref{fig:delta_mu_Co}, cobalt and nickel show similar features: the change in the absorption coefficients can be interpreted as the combination of a decrease in the photoabsorption in the narrower energy range around the $M_{2,3}$ absorption edge and an increase in the photoabsorption in the wider energy range. For cobalt, a significant decrease in the photoabsorption is observed even for a high electron temperature ($T_e=0.5$~eV). By contrast, for nickel, a relatively weak decrease in the absorption is observed, and this decrease is even overcome by the increase in the absorption observed for  high electron temperatures. This indicates that the narrow-energy-range decrease in the photoabsorption around the $M_{2,3}$ edge reflects a magnetic property of materials since cobalt has a larger magnetization than nickel.

\begin{figure}[htbp]
  \centering
  \includegraphics[width=0.95\columnwidth]{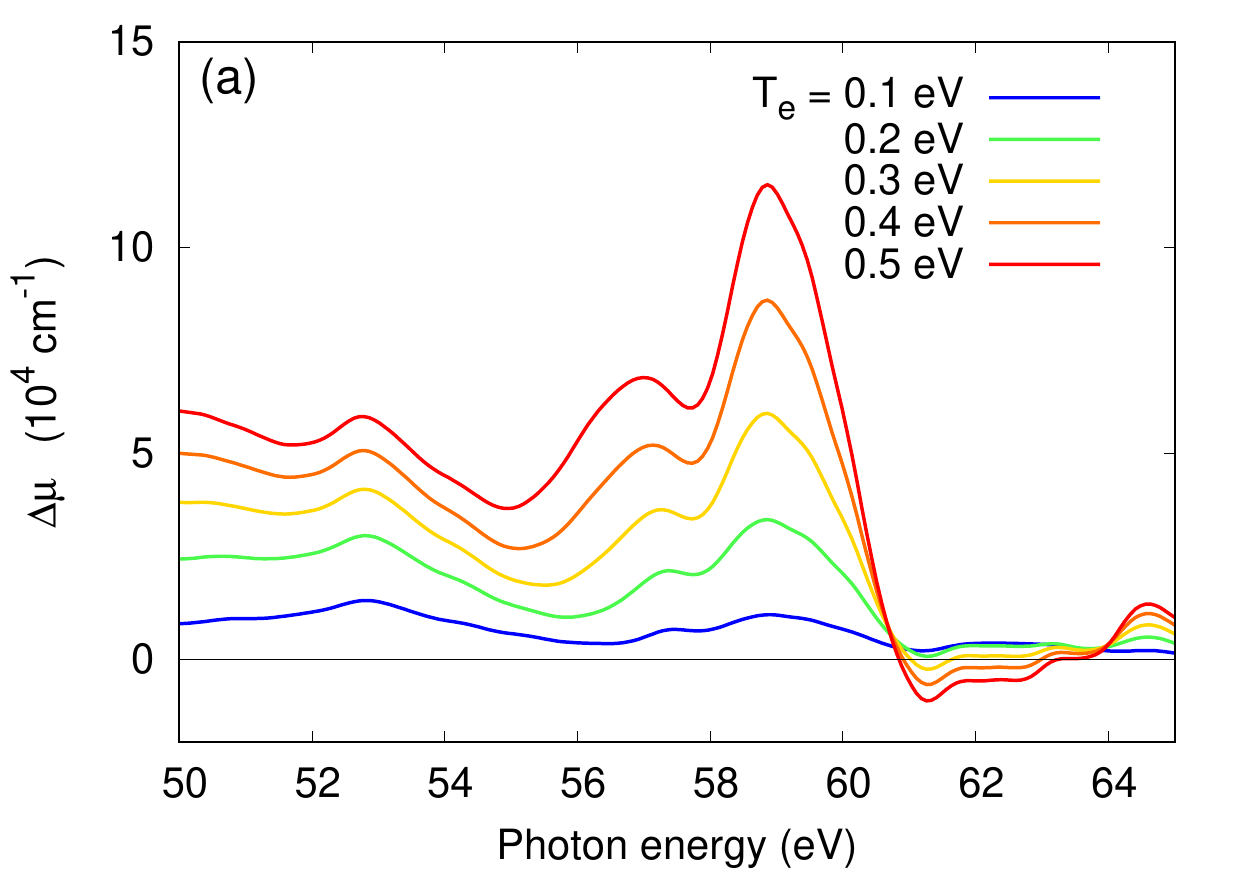}
  \includegraphics[width=0.95\columnwidth]{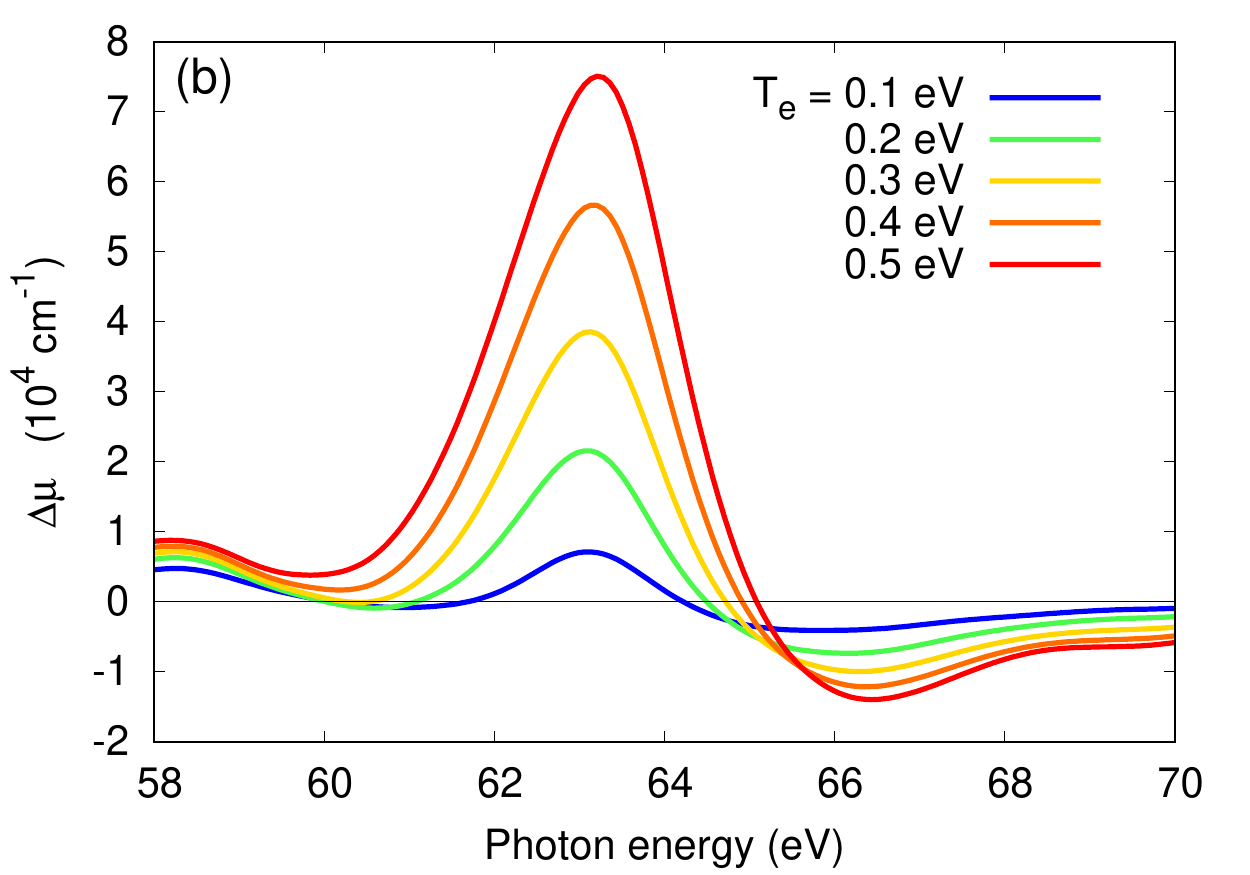}
\caption{\label{fig:delta_mu_Co_spinless}
Change in the photoabsorption, $\Delta \mu$, with the increase in the electron temperature, $T_e$, for (a) Co and (b) Ni. The results were obtained using the TDDFT with the spin-degenerate electronic configuration.
}
\end{figure}

To confirm this hypothesis, we evaluate the change in the absorption spectrum using the spin-degenerate electronic configuration. Figure~\ref{fig:delta_mu_Co_spinless} shows the change in the absorption coefficients with the increase in the electron temperature, $T_e$, for (a)~cobalt and (b)~nickel using the spin-degenerate electronic configuration. In agreement with our hypothesis, neither cobalt nor nickel show a significant decrease in the photoabsorption around the $M_{2,3}$ edge when the two spin components are degenerate. Therefore, we can confirm that the decrease n the photoabsrotion in the narrow energy range around the absorption edge in Figs.~\ref{fig:delta_mu_Co}~(a) and (b) is an intrinsic property of magnetic materials. As seen from Fig.~\ref{fig:delta_mu_Co}, both cobalt and nickel with the spin-degenerate configuration show a photoabsorption increase in the wide energy range around the absorption edge. A similar behavior has been observed in a previous experiment for laser-excited bulk titanium, and the photoabsortion increase has been explained in terms of the modification in the local-field effect due to light-induced ultrafast electron localization around transition metal elements \cite{Volkov2019}. Therefore, the wide-energy-range absorption increase in cobalt and nickel can be also understood as a general feature of transition metals due to the light-induced electron localization.

\begin{figure}[htbp]
  \centering
  \includegraphics[width=0.95\columnwidth]{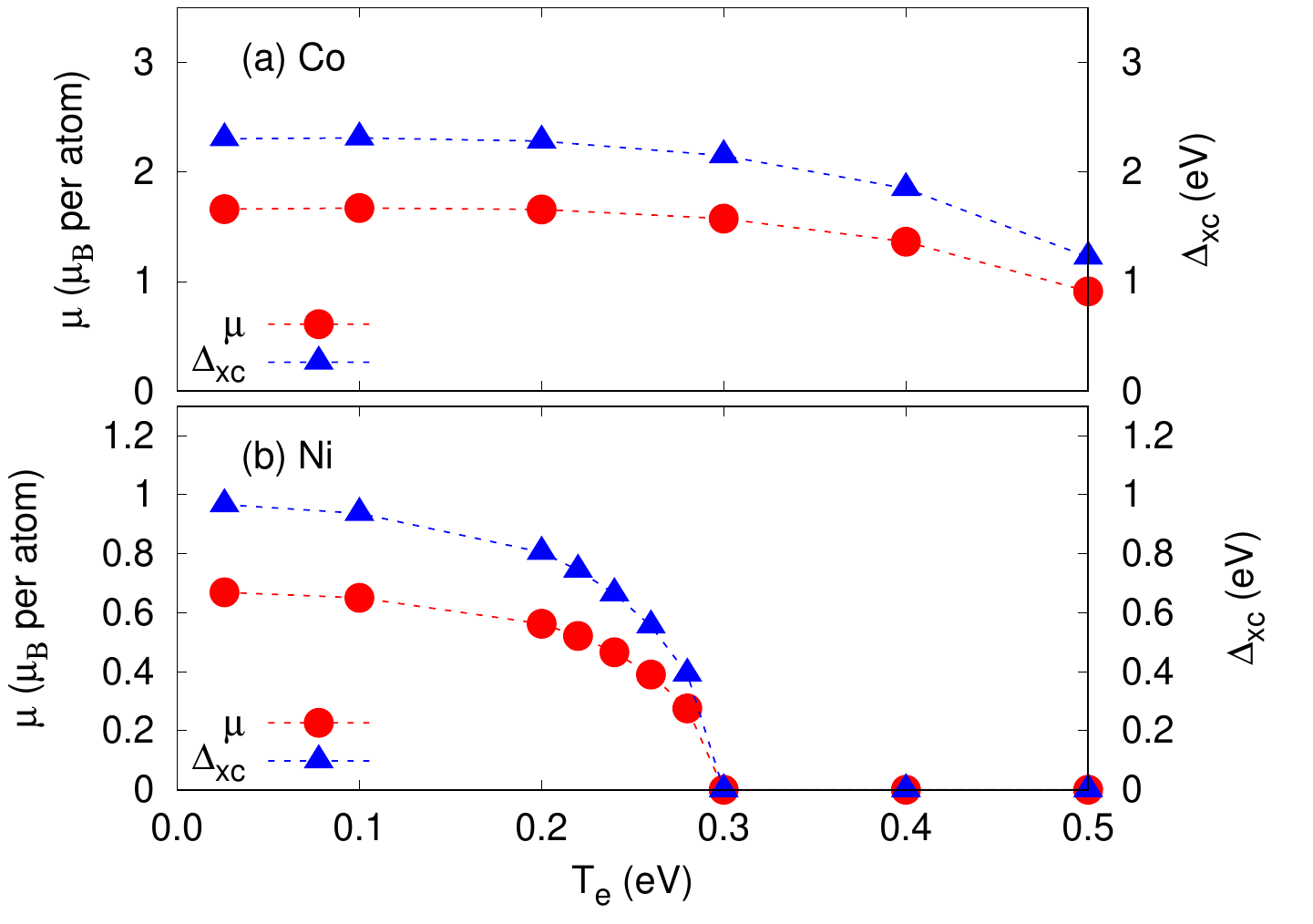}
\caption{\label{fig:mu_delta_xc_vs_Te}
Magnetic properties of (a) Co  and (b) Ni at a finite electron temperature. In each panel, the magnetic moment $\mu$ and the exchange-splitting of the $3p$ bands are shown as a function of the electron temperature.
}
\end{figure}

To develop a microscopic understanding of the observed decrease in photoabsorption around the $M_{2,3}$ absorption edge, we evaluate the magnetic properties of cobalt and nickel at a finite electron temperature, $T_e$. Figure~\ref{fig:mu_delta_xc_vs_Te} shows the magnetization, $\mu$, and the exchange split of the $3p$ bands, $\Delta_{xc}$, for bulk (a)~cobalt and (b)~nickel. As seen from Fig.~\ref{fig:mu_delta_xc_vs_Te}~(a), the magnetization of cobalt decreases with the increase in the electron temperature but remains finite at $T_e=0.5$~eV. By contrast, as seen from Fig.~\ref{fig:mu_delta_xc_vs_Te}~(b), the magnetization of nickel decreases with the increase in $T_e$ and disappears at $T_e=0.3$~eV. In both cases, the exchange split, $\Delta_{xc}$, shows a very similar trend to that of the magnetization. Furthermore, the evolution of the exchange split, $\Delta_{xc}$, upon increasing the electron temperature is similar to that observed for the photoabsorption: The photoabsorption of cobalt continues to decrease around the absorption edge for $T_e<0.5$~eV (see Fig.~\ref{fig:delta_mu_Co}~(a)), whereas that of nickel is saturated at around $T_e=0.3$~eV (see Fig.~\ref{fig:delta_mu_Co}~(b)) due to the closing of the exchange splitting of the nickel $3p$ bands, $\Delta_{xc}$. This indicates that the decrease in the photoabsorption around the $M_{2,3}$ absorption edge of laser-excited cobalt and nickel can be understood in terms of the blue shift of the absorption edge through the reduction in the exchange split upon increasing the electron temperature.

Based on the above analysis, we find that the change in the photoabsorption of laser-excited cobalt and nickel consists of the photoabsorption increase in a wider energy range around the absorption edge and the photoabsorption decrease in a narrow energy range around the absorption edge. The photoabsorption increase is a common feature of laser-excited transition metals, which reflects the laser-induced electron localization and modification in the microscopic screening properties. On the other hand, the photoabsorption decrease is a unique feature of laser-excited magnetic materials, which reflects the light-induced local demagnetization around the ions and the subsequent reduction in the exchange splitting of semicore states. Therefore, this finding indicates that element-specific local magnetization dynamics can be investigated via time-resolved transient absorption spectroscopy as transient magnetic properties are recoded in transient absorption spectra.

\section{Summary and Outlook \label{sec:summary}}

We studied the optical properties of laser-excited bulk cobalt and nickel by extending the first-principles calculation on the transient absorption spectroscopy with the TDDFT \cite{PhysRevB.90.174303} to magnetic materials. The simulation results show that the transient absorption of laser-excited magnetic materials mainly consists of two components. One is the decrease in the photoabsorption in the narrow energy range around the $M_{2,3}$ absorption edge, and the other is the increase in the photoabsorption in the wider energy range. The wider-range photoabsorption increase is consistent with a previous result reported for laser-excited titanium \cite{Volkov2019}, where the absorption increase had been explained in terms of the modification in the local-field effects through laser-induced electron localization. Therefore, the increase in the photoabsorption of cobalt and nickel upon increasing the electron temperature can be interpreted as a general feature of laser-excited transition metals. By contrast, we found that the photoabsorption decrease around the $M_{2,3}$ edge is an intrinsic property of magnetic materials. Based on the microscopic analysis of the magnetic properties at a finite electron temperature, the decrease in photoabsorption can be understood as the blue shift of the absorption edge due to the decrease in the exchange splitting.

The above analysis clarifies the relation between element-specific local magnetic properties and optical responses of semicore states of the corresponding element. Based on this finding, one may investigate light-induced magnetization and spin dynamics in the time domain using transient absorption spectroscopy. For example, the application of the attosecond transient absorption spectroscopy \cite{doi:10.1126/science.1260311,doi:10.1126/science.aag1268} to magnetic materials may enable access to the ultrafast spin dynamics and magnetization with the attosecond resolution. Therefore, in addition to the recently developed attosecond circular dichroism technique with circularly polarized probe pulses \cite{Siegrist2019}, the transient absorption spectroscopy with linearly polarized probe pulses can be a complementary and alternative approach to study real-time magnetization and spin dynamics in magnetic materials.

\begin{acknowledgments}
This work was supported by JSPS KAKENHI Grant Number JP20K14382. The author thanks Enago for the English language review.
\end{acknowledgments}
\bibliography{ref}

\end{document}